\documentclass[twocolumn,letterpaper,amsmath,amssymb,floatfix,prl,superscriptaddress]{revtex4} 

\usepackage{graphicx}
\usepackage{dcolumn}  
\usepackage{bm}  
\usepackage{epsfig}

\begin{document}
\setlength\arraycolsep{2pt}

\title{Thermal Casimir drag in fluctuating classical fields }

\author{Vincent D\'emery}
\affiliation{Universit\'e de Toulouse, UPS, CNRS, Laboratoire de Physique Th\'eorique (IRSAMC),  F-31062 Toulouse, France}

\author{David S. Dean}
\affiliation{Universit\'e de Toulouse, UPS, CNRS, Laboratoire de Physique Th\'eorique (IRSAMC),  F-31062 Toulouse, France}

\begin{abstract}
A uniformly moving inclusion which locally suppresses the fluctuations of a 
{\em classical} thermally excited field is shown to experience a  drag force which depends on the dynamics of the field. It is shown that in a number of cases the linear friction coefficient is dominated by short distance fluctuations and takes a very simple form. Examples where this drag can occur are for stiff objects, such as proteins, nonspecifically bound to more flexible ones such as polymers and membranes. 

\end{abstract}

\maketitle
The Casimir force, both quantum and thermal (or pseudo), arises due to the imposition of boundary conditions on  quantum or thermal fields \cite{kar1999}, the influence  of the field is manifested by the force induced between two or more particles or surfaces in the field. However the presence of the field can also be seen by looking at the force exerted on a {\em single} particle when it is not at rest, for example a quantum friction exists for a neutral atom moving parallel to a dielectric surface \cite{ann1986} and for parallel surfaces in relative motion \cite{pen2010}. While the latter frictional forces require the presence of a dielectric or conducting surface, it has also  been proposed that a  frictional Casimir force can be  induced by the uniform motion of a polarizable molecule in a volume of blackbody of radiation which is in equilibrium in the rest frame of a containing cavity \cite{mkr2003}. Frictional forces acting on impurities in superfluids and Bose Einstein condensates have also been studied, at low speeds no frictional force is found at a mean field level below a critical velocity \cite{ast2004}, however the scattering of quantum and thermal fluctuations can be shown to generate a friction at arbitrarily small velocities \cite{ast2004,rob2005,syk2009}.  

\begin{figure}
\epsfxsize=0.8\hsize \epsfbox{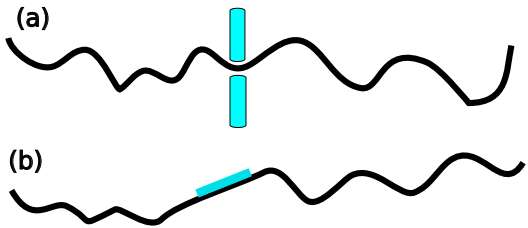}
\caption{Effects of  inclusion interactions (thicker line) visualized on a one dimensional 
polymer (thinner  line) (a) height constraint where the height of the polymer  is fixed at zero to pass
through a pore (b) a non flexible inclusion bound to the polymer which suppresses the local curvature.}
\label{fig1}
\end{figure} 

For inclusions linearly coupled to classical fields driven by thermal fluctuations (for example a protein imposing a local curvature in a lipid membrane or a colloid in a binary liquid which is preferentially wetted by one of the phases) we recently showed that  the insertion experiences a drag force when it moves at constant velocity \cite{demery2010}. The force arises because the insertion causes a local polarization of the fluctuating field, for instance a local magnetization for point-like field moving through a ferromagnet or a local non-zero curvature for a  curvature inducing protein in a membrane. This local polarization is not symmetric about the particle when the particle moves at constant velocity and an effective frictional force is induced which opposes the inclusion's motion. At low velocity the force is typically linear in the velocity, $f= -\lambda v$ where $\lambda$ defines an effective friction coefficient and at large velocities the force decays as $|f|\sim 1/v$. The average value of the force is in fact independent of the fluctuations of the field, it is mean field like and induced by the local symmetry breaking of the field by the inclusion.  An  electrodynamical analogy for this linear coupling case is the  drag on charged particles which create an average polarization (the polaron)  in the surrounding medium.  However a neutral atom does not induce an average polarization, but the fluctuations of the
field induced by the dipole interact with the fluctuations of the field in the surrounding medium
and it is this that leads to a frictional force. The {\em soft matter} analogue of this is  drag experienced by an insertion in a classical field, driven by thermal fluctuations, which {\em does not} break the local symmetry of the field.  In physical terms  examples would be colloids in binary mixtures which are not preferentially wetted or membrane inclusion which flatten out membrane fluctuations, due to their stiffness, rather than inducing a local mean curvature. We show that a drag force is present for a variety of free fields undergoing stochastic dissipative dynamics of a type often used to model dynamics in soft condensed matter systems.

The model we consider is for a scalar field with Hamiltonian
\begin{equation}
H[\phi,{\bf z}] = {1\over 2}\int d{\bf x} \ \phi({\bf x})\Delta\phi({\bf x})
+ \Delta H(\phi({\bf z}))
\end{equation}
where $\Delta$ is a positive definite operator which is model dependent, for instance
for $\Delta({\bf x}-{\bf y})  = [-\nabla^2 + m^2] \delta({\bf x}-{\bf y})$ the field $\phi$ can represent the
local magnetization in the Gaussian (high-temperature) approximation for the field theory of
a ferromagnet \cite{chai2000}. Similar models can be used to represent the height fluctuations of fluid-fluid/fluid-air interfaces \cite{ben} and lipid membranes  \cite{hel1973} and these models include operators with higher order derivatives due to the contribution of bending energy.  The second term $\Delta H$ represents the energy of interaction between the field and an insertion at the point ${\bf z}$. In this paper we will consider interactions that are quadratic in the field variable $\phi$, the interaction with the insertion  therefore does not affect the mean value of the field $\phi$ about the inclusion (which is zero) but  however modifies the field fluctuations. In this paper we will consider three interactions types
\begin{equation}
\Delta H= (a) \ {h\over 2}\phi^2({\bf z})\  (b) \  {h\over 2}[\nabla \phi({\bf z})]^2 \ 
(c)   {h\over 2}[\nabla^2 \phi({\bf z})]^2
\label{ints}
\end{equation}
For interaction (a) in the limit where $h\to\infty$ the insertion field 
coupling induces a point-like Dirichlet boundary condition where $\phi$ is energetically constrained
to be zero at the inclusion's position $\bf z$, an example of how this can occur is shown in Fig (1a), a region of a polymer is pinned at a fixed height $0$, this could be due to an optical trap
or due to a pore that the polymer passes through.  In the case (b) the gradient of the field is suppressed and in the limit $h\to \infty$ point like Neumann boundary conditions are imposed. Physically this corresponds to a region where the surface energy per unit area is large and the membrane is thus locally flattened, this type of interaction also occurs in the nematic phase of liquid
crystals where inclusions suppress or enhance fluctuations of the director field about its average orientation \cite{bar00}. Shown is Fig (1b) is the case (c), where stiff membrane inclusions suppress the local membrane curvature.   

We will study the force felt by the insertion when it moves at a constant velocity in the field. The dynamics of the field will be of a generic stochastic dissipative type 
\begin{equation}
{\partial \phi({\bf x})\over \partial t}= -R{\delta H\over \delta\phi({\bf x})}+\eta({\bf x},t)
\end{equation}
where the thermal noise driving the field is white in time and has a spatial correlation function obeying
the local fluctuation dissipation relation
\begin{equation}
\langle \eta({\bf x},t)\eta({\bf y},t)\rangle = 2T\delta(t-t')R({\bf x},{\bf y}),
\end{equation}
where $T$ is the temperature. Above we use the operator notation $AB({\bf x},{\bf y}) =
\int d{\bf x}' A({\bf x},{\bf x}') B({\bf x}',{\bf y})$, and we consider dynamical operators $R$ which
are invariant with respect to spatial translation, {\em i.e.} $R({\bf x},{\bf y}) = R(\bf{x}-{\bf y})$.

We outline the computation of the force acting on the particle for Dirichlet type insertion/field 
interactions of type Eq. (\ref{ints}a). The force on the particle is  given by
\begin{equation}
{\bf f} =-\nabla_{\bf z} H = -h\phi({\bf z})\nabla_{\bf z}\phi({\bf z}),
\end{equation}
where ${\bf z}= {\bf v}t$. The imposition of effective boundary terms via a local interaction with the field has the advantage that the expression for the local force can be unambiguously written down as above for any field configuration \cite{dego}. 

The average value of the force may be written in terms of the equal time correlation function of the field  $\langle \phi({\bf x},t)\phi({\bf y},t)\rangle =C({\bf x},{\bf y},t)$ as 
\begin{equation}
\langle {\bf f}\rangle = -h\ \nabla_{\bf z}C({\bf z}, {\bf z}')\vert_{{\bf z}={\bf z}'={\bf v}t}.
\end{equation}
In terms of the comoving coordinates of the insertion the correlation function can be shown to obey
\begin{eqnarray}
&&{\partial \over \partial t }C({\bf x}',{\bf y}',t)-({\bf v}\cdot\nabla_{{\bf x}'}+{\bf v}\cdot\nabla_{{\bf y}'})C({\bf x}',{\bf y}',t) \nonumber \\
&=&-(R\Delta_{{\bf x}'}+ R\Delta_{{\bf y}'})C({\bf x}',{\bf y}',t)-hR_{{\bf x}'}C({\bf x}',{\bf y}',t)\delta({\bf x}')\nonumber \\
&&- hR_{{\bf y}'}C({\bf x}',{\bf y}',t)\delta({\bf y}')
+2TR({\bf x}'-{\bf y}'),
\end{eqnarray}
where $A_{\bf x} C({\bf x},{\bf y})=\int d{\bf x}' A({\bf x},{\bf x}') C({\bf x}',{\bf y}) $ denotes operation on the (first) argument ${\bf x}$ of the correlation function $C$. 
In the steady state regime the double Fourier transform of the  correlation function obeys
\begin{eqnarray}
&& \left[\tilde R({\bf p})\tilde\Delta({\bf p})+\tilde R({\bf q})\tilde\Delta({\bf q})
-i{\bf v}\cdot({\bf p}+{\bf q}) \right]\tilde C({\bf p},{\bf q}) \nonumber \\
&+&h\tilde R({\bf p})\int {d{\bf q}'\over (2\pi)^{d}}\tilde C({\bf q}',{\bf q})+h\tilde R({\bf q})\int {d{\bf q}'\over (2\pi)^{d}}\tilde C({\bf q}',{\bf p}) \nonumber \\
&=& 2T(2\pi)^d\tilde R({\bf p})\delta({\bf p}+{\bf q}).
\end{eqnarray}
In this notation the average force can be written as 
\begin{equation}
\langle {\bf f}\rangle = -ih\int { d{\bf q}\over (2\pi)^{d}}\; {\bf q}\tilde A({\bf q}),
\end{equation}
where
\begin{equation}
A({\bf q}) = \int {d{\bf q}'\over (2\pi)^{d}}\tilde C({\bf q}',{\bf q}),
\end{equation}
and can be shown to obey
\begin{eqnarray}
A({\bf q}) &+& h\int {d{\bf p}\over (2\pi)^d} {\tilde R({\bf p})A({\bf q})+\tilde R({\bf q})A({\bf p})\over 
\tilde R({\bf p}) \tilde\Delta({\bf p})+\tilde R({\bf q})\tilde\Delta({\bf q})
-i{\bf v}\cdot({\bf p}+{\bf q})}\nonumber\\  &=& {T\over \tilde \Delta({\bf q})}.\label{eqA}
\end{eqnarray}

We can solve Eq. (\ref{eqA}) in two limiting cases to give a general picture of how the 
frictional force behaves.

\noindent{\em  (i) Weak interaction limit:} In the case where $h$ is small we can compute 
$A$ via a perturbation expansion in $h$, writing $A=A_0 + hA_1 + O(h^2)$ we find that
\begin{equation}
A_0({\bf q}) = {T\over \tilde \Delta({\bf q})}
\end{equation}
and 
\begin{eqnarray}
&&A_1({\bf q}) = -T \int {d{\bf p}\over (2\pi)^d}\nonumber \\&& {\tilde R({\bf p})\tilde \Delta({\bf p})+\tilde R({\bf q})\tilde \Delta({\bf q})\over \tilde \Delta({\bf q})\tilde \Delta({\bf p})\left[ \tilde R({\bf p})\tilde \Delta({\bf p})+\tilde R({\bf q})\tilde \Delta({\bf q})-i{\bf v}\cdot({\bf p}+{\bf q})\right]}\nonumber \\
\end{eqnarray}
To this lowest order  the force is  given by
\begin{eqnarray}
&& \langle {\bf f}\rangle =-{Th^2\over 2}\int {d{\bf p}d{\bf q}\over (2\pi)^{2d}}\nonumber \\ 
&&{a({\bf p},{\bf q})({\bf p}+{\bf q})\ {\bf v}\cdot({\bf p}+{\bf q})\left(\tilde R({\bf p})\tilde \Delta({\bf p})+\tilde R({\bf q})\tilde \Delta({\bf q})\right)\over \tilde \Delta({\bf q})\tilde \Delta({\bf p})\left(\left[ \tilde R({\bf p})\tilde \Delta({\bf p})+\tilde R({\bf q})\tilde \Delta({\bf q})\right]^2 + \left[{\bf v}\cdot({\bf p}+{\bf q})\right]^2\right)}\nonumber \\
\label{genpert}
\end{eqnarray}
with $a({\bf p},{\bf q}) = 1$. The equivalent of Eq. (\ref{genpert}) can also be derived for the case
of the Neumann type interaction -- Eq. (\ref{ints}b) -- and curvature suppressing interactions
-- Eq. (\ref{ints}c) -- and the corresponding results have $a({\bf p},{\bf q}) = ({\bf p}\cdot{\bf q})^2$
and $a({\bf p},{\bf q}) = p^4 q^4$ respectively. In the above expressions the Fourier integrals may diverge and they are naturally regularized by 
an ultra-violet cut-off $\Lambda = \pi/a$ where $a$ corresponds to a microscopic size below which the
field cannot fluctuate. In principal the cut-off could also be the size of the inclusion, however for simplicity
in this study we consider inclusions of the same size as the microscopic cut-off. Here it is clear that 
for small $v$ the force is generically linear in $v$ but for large $v$ we have
$|f|\sim 1/v$ as in the case of linear couplings to the field \cite{demery2010}. Finally we see that the
force is fluctuation induced as it is proportional to $T$ and is a second order effect in the coupling $h$.

\noindent{\em The small velocity regime:} 
When $v=0$ we find that 
\begin{equation}
A_0({\bf q}) = {T\over \tilde \Delta({\bf q})}{1\over 1 + h\int {d{\bf q}'\over (2\pi)^d}{1\over \tilde \Delta({\bf q}')}}.
\end{equation}
For small $v$ we assume that $A({\bf q})$ has a correction term $A_1({\bf q})$ which is of order $v$:
\begin{equation}
A({\bf q}) = A_0({\bf q}) + A_1({\bf q}) + O(v^2).
\end{equation}
from which we find that $A_1$ is given by
\begin{equation}
A_1({\bf q}) = -ih{\bf v}\cdot{\bf q}{ \int {d{\bf p}\over (2\pi)^d} {\tilde R({\bf p})A_0({\bf q})
+\tilde R({\bf q})A_0({\bf p})\over \left[ \tilde R({\bf p})\tilde\Delta({\bf p}) + \tilde R({\bf q})\tilde \Delta({\bf q})
\right]^2}\over 
1 + h\int{d{\bf p}\over (2\pi)^d}{R({\bf p})\over \tilde R({\bf p})\tilde \Delta({\bf p}) + \tilde R({\bf q})
\tilde \Delta ({\bf q})}}.
\end{equation}
For simplicity we consider the limit of a Dirichlet point boundary condition where $h\to\infty$ 
and we find the result:
\begin{equation}
 \lambda ={cT\over  d \int {d{\bf q}'}
{q^{\alpha}\over \tilde \Delta({\bf q}')}}  \int d{\bf q} q^{2+\alpha} { \int {d{\bf p}p^{\alpha}\over \tilde \Delta({\bf q})\tilde \Delta({\bf p})\left[ \tilde R({\bf p})\tilde\Delta({\bf p}) + \tilde R({\bf q})\tilde \Delta({\bf q})
\right]}\over 
\int d{\bf p}p^{\alpha}{\tilde R({\bf p})\over \tilde R({\bf p})\tilde \Delta({\bf p}) + \tilde R({\bf q})
\tilde \Delta ({\bf q})}},\label{ld}
\end{equation}
for the friction coefficient, with $c=1$ and $\alpha =1$. In the same limit  $h\to\infty$  we find a result of the same form with $c=d$ and $\alpha = 2$ for the Neumann  case and  $c=1$ and $\alpha =4$ for the curvature suppressing case.

As the operators $\Delta$ and $R$ are isotropic, their Fourier transforms are functions
of $q=|{\bf q}|$, from Eq. (\ref{ld}) and the forms for $\alpha$ given after, we find the  result that all of these results are related as a  function of the spatial dimension $d$ via 
\begin{equation}
\lambda_N(d)=\lambda_D(d+2)d\ {\rm and}\ \lambda_C(d)=\lambda_D(d+4),\label{dnr}
\end{equation}
the subscripts $D$ $N$ and $C$ denoting Dirichlet, Neumann and Curvature suppressing. 

{\noindent \em Model A ferromagnet-} Here we consider a simple fluctuating field where
$\tilde \Delta(q) = q^2 + m^2$ and we take simple diffusive dynamics for the field, characterized
by a diffusion constant $D_0$ so that $\tilde R(q) = D_0$. For $d\geq 3$ for Dirichlet and 
$d\geq 1$ in the Neumann case and all D for the curvature suppressing case,   the friction coefficient takes a simple form in the limit where $ma\to 0$, i.e. where the microscopic cut-off $a$ is much smaller than the correlation length $\xi=1/m$ of
the field
\begin{equation}
\lambda_D(d) = {T\over D_0} Q(d) \label{defq}
 \end{equation}
 where $Q(d)$ can be expressed as integrals independent of $a$ and $m$. For example
$Q(3) \approx 0.479$ in the case of a sharp cut-off as used here. However $Q(d)$ will generally depend precisely on how the theory is regularized at short distances. In the cases $d=1$ and $d=2$  we find that the friction coefficient takes the form $\lambda_D = Tf(ma)/D_0 $ and where $f(x)=-2\ln(x)/\pi$ and $f(x) = \ln(-\ln(x))/2$ respectively.

{\noindent \em Model B ferromagnet-} Here we again take the Gaussian ferromagnet but consider the case of model B dynamics which conserves the total magnetization and where $\tilde R(q) = D_0 a^2 q^2$, where $D_0$ is again a diffusion constant and the dynamical operator is given the correct
physical dimensions via its dependence on the cut-off length scale $a$. In this notation we find for
the Dirichlet interaction that for $d\geq 4$ $\lambda_D$ takes the form of Eq. (\ref{defq}) (the dependence on the cut-off $a$ cancels with the factor of $a^2$ appearing in the definition of $\tilde R$ and ultra-violet divergences in the integrals in the expression for $\lambda$). For $d=3$ we find that $\lambda
\sim -T\ln(ma)/D_0$, $\lambda \sim T/[D_0(ma \ln(ma))^2]$ for $d=2$
 and for $d=1$ we have $\lambda\sim T/[D_0(ma)^2]$.
 
 {\noindent \em Rigid inclusion in lipid bilayers-} Here we consider the case of a rigid inclusion in a lipid bilayer which locally suppresses the membrane curvature. The interaction term
is of the form Eq. (\ref{ints}c), and using the Helfrich hamiltonian \cite{hel1973} gives 
$\tilde\Delta(q) = \kappa q^4 +\sigma q^2$, where $\kappa$ and $\sigma$ are respectively the
membrane bending rigidity and surface tension. Membrane fluctuation dynamics  is dominated by the
surrounding fluid and we have $\tilde R(q)=1/4\eta q$, where $\eta$ is the viscosity of the surrounding fluid \cite{lin2004}. Here the drag is dominated by short distance behavior and we find the result $\lambda_C\sim T\eta a/\kappa$, in the limit where $a\ll \xi=\sqrt{\kappa/ \sigma}$, (the correlation length of the membrane fluctuations). 

\begin{figure}
\epsfxsize=\hsize \epsfbox{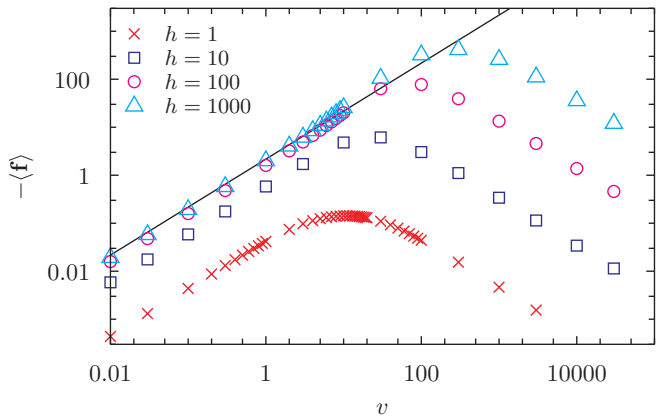}
\caption{Drag force vs. velocity for a model A ferromagnet in $d=1$ (on a log-log scale),
with different values of $h$. The solid line is the analytical result
$\lambda_D v$.}
\label{fig2}
\end{figure}

Eq. (\ref{eqA}) can also be solved numerically for any $h$ and $v$. We
performed this numerical computation for a model A ferromagnet in $d=1$; with a 
Dirichlet inclusion/field interaction and  setting  $ma=0.1$,
The computation was done with 101 Fourier modes. The mean drag force
as a function of the velocity is plotted in Fig. (\ref{fig2}) for
different values of $h$. We see that for small $v$ $\langle f\rangle$ is linear in $v$ (and for
large $h$ the coefficient agrees with our analytical result) and
that for large $v$ it decays as $1/v$. It is also interesting to note that the regime in $v$ where linear frictional forces arise is very large, and its range increases upon increasing the interaction strength $h$.

We have seen that in a number of cases, in particular when the insertion coupling depends on higher
derivatives of the field and when the spatial dimension is high, the friction is dominated by the short wave length fluctuations of the field. In these cases it can be shown from Eq. (\ref{ld}) (and 
dimensional analysis)  that 
\begin{equation}
\lambda \sim T {\Lambda^2 \tau(\Lambda)}
\end{equation}
where $\tau(\Lambda)= [\tilde R(\Lambda)\tilde \Delta(\Lambda)]^{-1}$ is the relaxation time of the shortest wave length. The scaling of all the short-distance dominated results above can  be deduced from this formula. The frictional force in this case is thus increased as the relaxation time of the short wave length modes is increased. The dependence of the friction coefficient on the microscopic cut-off is rather subtle. In the low dimensional ferromagnets with model A and B dynamics the friction coefficient increases as the cut-off decreases, but for  curvature suppressing proteins in a Helfrich model membrane driven by stochastic hydrodynamics, the drag increases as the cut-off
size increases.  

In the study of \cite{mkr2003} of the drag on a polarizable  molecule in blackbody radiation, one can see that the classical contribution to this drag is also proportional to $T$, as found in our study. However in \cite{mkr2003}, the drag is of first order in the molecule's polarizability, whereas the effect found here is second order in the parameter $h$ characterizing the interaction strength between the field and the particle.  The drag experienced by impurities in one-dimensional Bose Einstein condensates \cite{syk2009} shows a similar behavior to our result, in the high temperature regime the force is proportional to the temperature and perturbatively the force is second order in the coupling $h$ of the impurity to the condensate. Physically the mechanism for drag creation is quite
different in the dissipative systems studied here, in the models presented here the inclusion is repelled
from regions where the field fluctuates. The region behind a moving inclusion fluctuates less than that ahead as the passage of the inclusion has flattened out the fluctuations. The field behind the inclusion is thus {\em calmer} than that ahead, thus making it energetically favorable for the inclusion to move backwards and thereby inducing the drag force.

The experimental observation of this drag requires a system where the fluctuation induced drag is
not swamped by other drags, such as hydrodynamic Stokes' drag. Our result for drag on a curvature
suppressing membrane protein appears to be only $1\%$ of the hydrodynamic drag given by the Saffman Delbr\"uck formula \cite{saff1975} for small insertions of the order of the membrane thickness. This is because the former is dominated by the viscosity of the embedding fluid and the latter by the membrane viscosity. However the addition of glycerol to the external solution can increase its viscosity by a factor of up to 50 and in this regime the fluctuation induced drag will be of the same order as the hydrodynamic one and perhaps the drag predicted here could be observable.


\begin{thebibliography}{99}
\bibitem{kar1999} M. Kardar and R. Golestanian, Rev. Mod. Phys. {\bf 71}, 1233  (1999).
\bibitem{ann1986}J.F. Annett and P.M. Echenique, Phys. Rev. B {\bf 34} 6853 (1986); {\em ibid} Phys. Rev. B {\bf 36} 8986 (1987).
\bibitem{pen2010} J.B. Pendry, New J. Phys. {\bf 12} 033028 (2010).
\bibitem{mkr2003}V. Mkrtchian, V.A. Parsegian, R. Podgornik and W.M. Saslow, Phys. Rev. Lett.
{\bf 91}, 220801 (2003)
\bibitem{ast2004} G.E. Astrakharchik and L.P. Pitaevskii, Phys. Rev. A {\bf 70}, 013608 (2004).
\bibitem{rob2005} D.C. Roberts and Y. Pomeau, Phys. Rev, Lett. {\bf 95}, 145303 (2005).
\bibitem{syk2009}A.G. Sykes, M.J. Davis and D.C. Roberts, Phys. Rev. Lett. {\bf 103}, 085302 (2009).
\bibitem{demery2010} V. D\'emery and D.S. Dean,  Phys. Rev. Lett.
{\bf 104}, 080601, (2010); V. D\'emery and D.S. Dean, Eur. J. Phys E {\bf 32}, 377 (2010).
\bibitem{chai2000}P.M. Chaikin and T.C. Lubensky, Principles of Condensed Matter Physics (Cambridge University Press, Cambridge) (2000).
\bibitem{ben}J. Huang, B. Davidovitch, C.D. Santangelo, T.P. Russell, and N. Menon,
Phys. Rev. Lett. {\bf 105}, 038302 (2010). 
\bibitem {bar00}D. Bartolo, D. Long and J.-B. Fournier, Europhys. Lett. {\bf 49}, 729 (2000).
\bibitem{hel1973} W. Helfrich, Z. Naturforsch. {\bf 28}c, 693 (1973).
\bibitem{dego} D.S. Dean and A. Gopinathan,  J. Stat. Mech. L08001 (2009); D.S. Dean and A.J. Gopinathan,Phys. Rev. E {\bf 81}, 041126 (2010).
\bibitem{lin2004} L.C.-L. Lin and F.L.H. Brown, Phys. Rev. Lett. {\bf 93}, 256001 (2004).
\bibitem{saff1975}P.G. Saffmann and M. Delbr\"uck, Proc. Natl. Acad. Sci USA {\bf 72}, 3111 (1975).
\end{thebibliography}
\end{document}